\documentclass[authoryear,3p,times]{elsarticle}
\usepackage{natbib}
\usepackage{graphicx}
\usepackage{epsfig,subfigure}
\usepackage{amssymb}
\usepackage{amsthm}
\usepackage{amsmath}

\journal{New Astronomy}
\begin{document}
\begin{frontmatter}
\title{The first photometric analysis of the W-subtype contact binary UCAC4 436-062932 with O'Connell effect}

\author[1,2,3]{Zhou, X.}
\author[1,2,3]{Qian, S.-B.}
\author[4]{Essam, A.}
\author[1,2]{He, J.-J.}
\author[1,2,3]{Zhang, B.}

\address[1]{Yunnan Observatories, Chinese Academy of Sciences, PO Box 110, 650216 Kunming, China}
\address[2]{Key Laboratory of the Structure and Evolution of Celestial Objects, Chinese Academy of Sciences, PO Box 110, 650216 Kunming, China}
\address[3]{Graduate University of the Chinese Academy of Sciences, Yuquan Road 19, Sijingshang Block, 100049 Beijing, China}
\address[4]{National Research Institute of Astronomy and Geophysics, Department of Astronomy, Helwan, Cairo, Egypt}
\cortext[cer]{Corresponding author. Tel.: +8613658806295 \\
E-mail address: zhouxiaophy@ynao.ac.cn}


\begin{abstract}
Two sets of light curves in $V$ $R_c$ $I_c$ bands for a newly discovered binary system UCAC4 436-062932 are obtained and analyzed using the Wilson-Devinney (W-D) code. The two sets of light curves get almost consistent results. The determined mass ratio is about $q = 2.7$ and the less massive component is nearly $250K$ hotter than the more massive one. The solutions conclude that UCAC4 436-062932 is a W-subtype shallow contact (with a contact degree of $f = 20\,\%$) binary system. Since the O'Connell effect appears on one set of the light curves, theories proposed to explain the effect are discussed. We assume that spot model may be the more plausible one to the O'Connell effect appeared on the asymmetric light curves of the binary system UCAC4 436-062932. Therefore, we add a cool spot on the surface of the more massive star (component with lower effective temperature) and get a quite approving results for the light curve fitting. It will provide evidence to support the spot model in the explanatory mechanism of O'Connell effect.
\end{abstract}


\end{frontmatter}

\section{Introduction}
In 1951, \citet{1951PRCO....2...85O} found that the light curve of the eclipsing binary RX Cas differed in brightness at the maxima proceeding and following the principal minimum. It was suggested to be caused by tidal effects and an increase in mutual radiation at periastron between stars moving in eccentric orbits (the so-called periastron effect in close eclipsing binaries) by \citet{1906MNRAS..66..123R} long time ago. However, \citet{1951PRCO....2...85O} pointed out that the effect had no connection with orbital eccentricity but caused by the circumstellar dust and gas. Then, it was called as the O'Connell effect. The O'Connell effect commonly appeared on light curves of contact and near contact binaries, which becomes a challenging problem in modeling the eclipsing binary systems. Several theories have been proposed to explain it. \citet{1990ApJ...355..271Z} assumed that the different brightness between the maxima can be explained in terms of the asymmetry of circumfluence in the contact binary star due to the Coriolis force. \citet{2003ChJAA...3..142L} suggested that the circumstellar material of a binary
system captured by its components is the key factor to cause the O'Connell effect. A star spot model was proposed for W-type systems with asymmetric light curves by \citet{1977VA.....21..359B} for the first time, and cool spot or hot spot model was widely used in other binary systems since then.

UCAC4 436-062932 (=2MASS J15582807-0257536) is a newly discovered  \citep{2014IBVS.6200....2E} late-type W UMa contact binary. In the present work, two sets of $V$ $R_c$ $I_c$ bands light curves of UCAC4 436-062932 with and without O'Connell effect are observed and analyzed using the Wilson-Devinney (W-D) code. Basing on the photometric solutions, the basic properties of the binary system is discussed. We also focus on the origin and evolution of the O'Connell effect appears on the light curves to give some evidence which will support the spot model in O'Connell effect explanation.

\section{The CCD photometric light curve observations of UCAC4 436-062932}

The CCD photometric observations of the eclipsing binary UCAC4 436-062932 were obtained on 2014 May 23, June 18 and June 19 using an EEV CCD 42-40 camera attached to the Newtonian focus of the 1.88m Kottamia reflector telescope in Egypt \citep{2014IBVS.6200....2E}. The CCD 42-40 camera has a format of $2048\times2048$ pixels with a scale of $0.''305$ pixel$^{-1}$  that was cooled by liquid nitrogen to $-125^\circ$C. During the observation, the broadband Johnson-Cousins $V$ $R_c$ $I_c$ filters were used. PHOT (measured magnitudes for a list of stars) of the aperture photometry package in the IRAF \footnote {The Image Reduction and Analysis Facility is hosted by the National Optical Astronomy Observatories in Tucson, Arizona at URL iraf.noao.edu.} was used to reduce the observed images. UCAC4 436-062934 and UCAC4 436-062914 were chosen as the comparison and the check stars. The coordinates and brightness in $V$ band of the variable, the comparison and the check stars are listed in Table \ref{Coordinates}. One of the CCD image in the field of view around UCAC4 436-062932 are shown in Figure 1, where``V'' refers to the eclipsing binary system, ``C'' to the comparison star and ``Ch'' to the check star.

\begin{table}
\begin{center}
\caption{\scriptsize Coordinates of UCAC4 436-062932, the comparison, and the check stars.}\label{Coordinates}
\begin{tabular}{ccccc}\hline\hline
Targets          &   name               & $\alpha_{2000}$        &  $\delta_{2000}$         &  $V_{mag}$      \\ \hline
Variable         &   UCAC4 436-062932   &$15^{h}58^{m}28^{s}.1$  & $-02^\circ57'53''.6$     &  $15.02$         \\
The comparison   &   UCAC4 436-062934   &$15^{h}58^{m}29^{s}.6$  & $-02^\circ59'53''.9$     &  $13.99$          \\
The check        &   UCAC4 436-062914	&$15^{h}58^{m}05^{s}.3$  & $-02^\circ54'40''.2$     &  $12.36$        \\
\hline\hline
\end{tabular}
\end{center}
\end{table}

\begin{figure}[!h]
\begin{center}
\includegraphics[width=8cm]{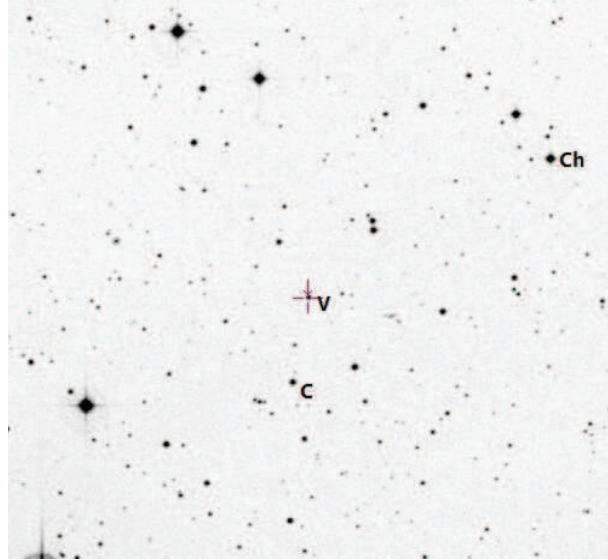}
\caption{CCD image in the field of view around UCAC4 436-062932 (V), the comparison star (C), and the check star (Ch).}
\end{center}
\end{figure}

The least- square method was used to determine the time of minimum light. The first linear ephemeris of UCAC4 436-062932 is obtained,
\begin{equation}
Min. I(HJD) = 2456827.4680(3) + 0^{d}.361456\times{E}.
\end{equation}
The light curves in $V$ $R_c$ $I_c$ bands are displayed in Figure 2 and Figure 3. As shown in Figure 2, the brightness of the two maxima in the light curves is almost equal. However, the light curves in Figure 3 show a negative O'Connell effect that the light maximum following the primary minimum is lower than another one.

\begin{figure}[!h]
\begin{center}
\includegraphics[width=12cm]{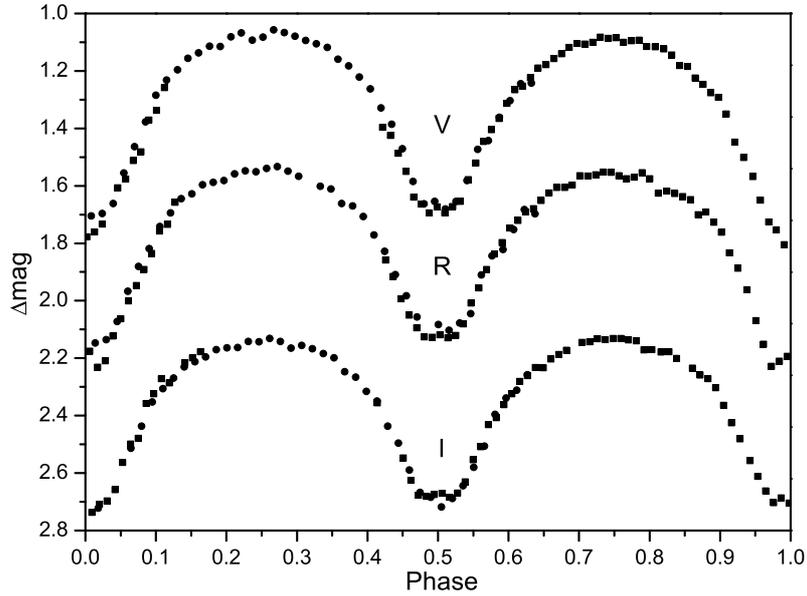}
\caption{CCD photometric light curves in $V$ $R_c$ and $I_c$ bands observed on 2014 May 23 and June 19. Circles and squares correspond to the data observed on May 23 and June 19, respectively. The brightness of the two maxima is almost equal.}
\end{center}
\end{figure}

\begin{figure}[!h]
\begin{center}
\includegraphics[width=12cm]{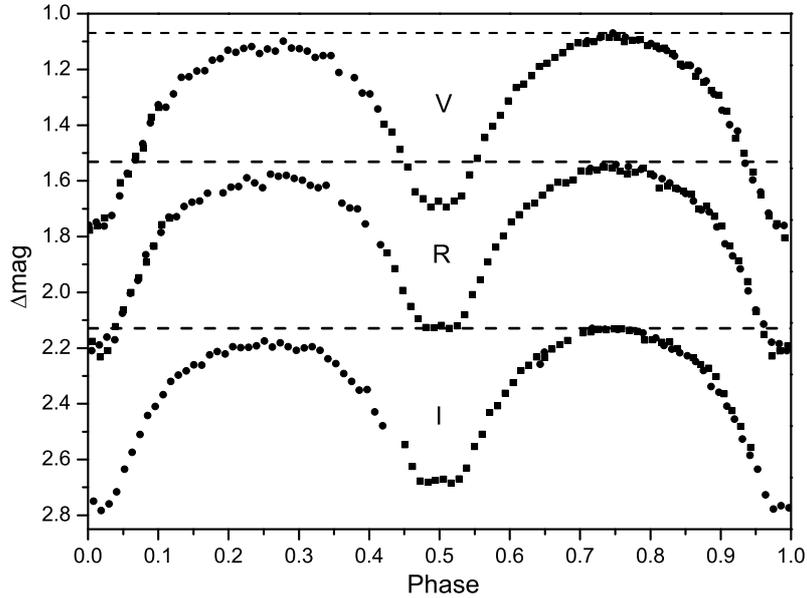}
\caption{CCD photometric light curves in $V$ $R_c$ and $I_c$ bands observed on 2014 June 18 and June 19. Circles and squares correspond to the data observed on June 18 and June 19, respectively. O'Connell effect appears on the light curves.}
\end{center}
\end{figure}

\section{Photometric solutions of UCAC4 436-062932}
UCAC4 436-062932 is a newly determined binary system and the light curves are typical EW type. The nearly flat eclipse reveals that UCAC4 436-062932 is a totally eclipsing binary system. In Figure 2 and Figure 3, the phases are calculated with Equation (1).
To understand its geometrical structure and evolutionary state, the $V$ $R_c$ and $I_c$ light curves shown in Figure 2 and Figure 3 are analyzed using W-D program of the 2013 version \citep{Wilson1971,Van2007,Wilson2010}.

Considering the fourth US Naval Observatory CCD Astrograph Catalog (UCAC4) gave the color index of $J - H =0.42$, we fixed the effective temperature of the primary star (star eclipsed at the primary minimum light) to be $T_1 = 5230K$ \citep{Cox2000}. It means that UCAC4 436-062932 is a late- type contact binary with convective envelopes. Therefore, the gravity-darkening coefficients $g_1=g_2=0.32$ \citep{1967ZA.....65...89L} and the bolometric albedo $A_1=A_2=0.5$ \citep{1969AcA....19..245R} are used. To account for the limb darkening in detail, logarithmic functions are used. The corresponding bolometric and passband-specific limb-darkening coefficients are chosen from \citet{1993AJ....106.2096V}'s table.

Since no spectroscopic results or photometric solution of UCAC4 436-062932 has been published, a $q$-search method is used to determine the initial input mass ratio for both of the two sets of light curves. During the calculating, it is found that the solution converges at mode 3 (contact model), and the adjustable parameters are: the mass ratio $q$ $(M_2/M_1)$; the orbital inclination $i$; the effective temperature of star 2 ($T_{2}$); the monochromatic luminosity of star 1 ($L_{1V}$, $L_{1R}$ and $L_{1I}$); the dimensionless potential of star 1 ($\Omega_{1}=\Omega_{2}$ in mode 3 for contact configuration). Solutions with mass ratio from 0.1 to 4.2 are investigated, and the relation between the resulting sum of weighted square deviations $\Sigma$ and $q$ is plotted in Figure 4. In the two sets of light curves, both of their minimum values are found at $q$ = 2.7, which indicates that UCAC4 436-062932 is a W-subtype \citep{1970VA.....12..217B} contact binary. Then $q$ = 2.7 is set as the initial value and considered as an adjustable parameter. The final photometric solutions for light curves without O'Connell effect are listed in Table 2 and the corresponding theoretical light curves are displayed in Figure 5. The results of light curves with O'Connell effect are listed in Table 3 and the corresponding theoretical light curves are displayed in Figure 6. It has to be mentioned that we model the light curves with O'Connell effect separately in different bands to get better results as a cool spot is added. The contact configuration (with spot) of UCAC4 436-062932 is displayed in Figure 7.

\begin{figure}[!h]
\begin{center}
\includegraphics[width=16cm]{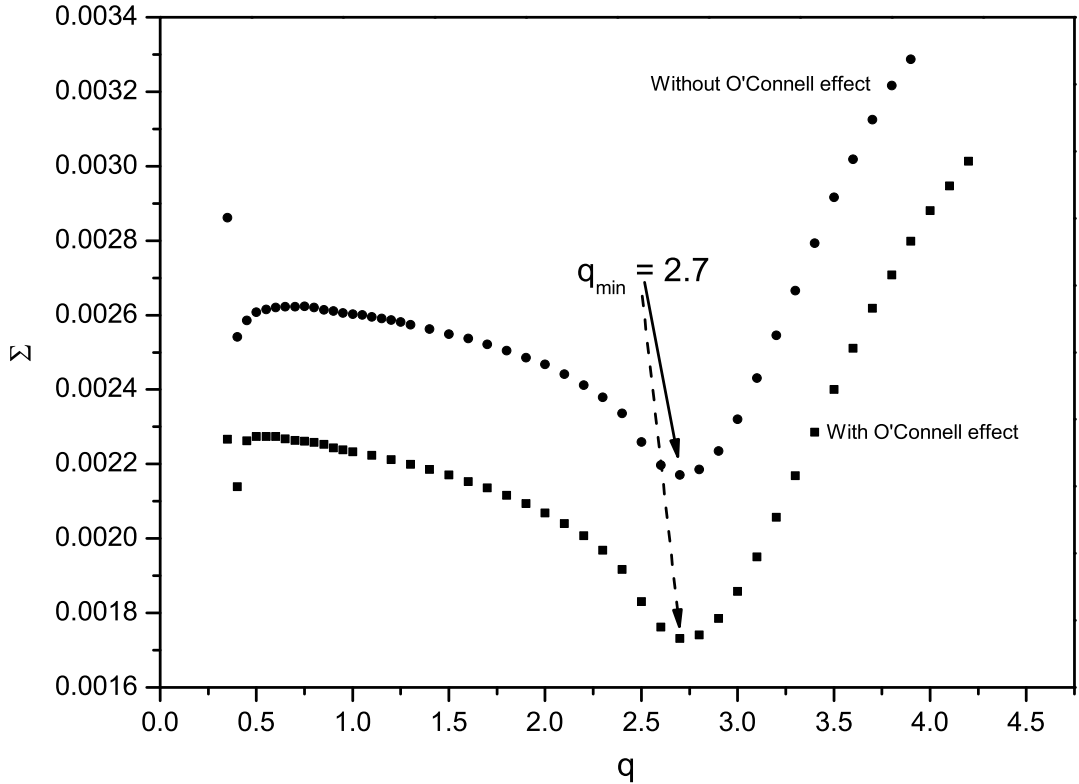}
\caption{Relation between $\Sigma$ and $q$ for light curves of UCAC4 436-062932. Circles represent the results without O'Connell effect and squares represent the results with O'Connell effect. $\Sigma$ is the resulting sum of weighted square deviations. It is shown that the minimum is at q = 2.70.}
\end{center}
\end{figure}

\begin{table}[!h]
\caption{Photometric solutions of UCAC4 436-062932 without O'Connell effect}\label{phsolutions}
\begin{center}
\begin{tabular}{lllllllll}
\hline
Parameters                        & Values                         \\
\hline
$g_{1}$                           & 0.32(fixed)                   \\
$g_{2}$                           & 0.32(fixed)                   \\
$A_{1}$                           & 0.50(fixed)                   \\
$A_{2}$                           & 0.50(fixed)                   \\
q ($M_2/M_1$ )                    & 2.72($\pm0.04$)             \\
$T_{1}(K)   $                     & 5230(fixed)                   \\
$i(^{\circ})$                     & 84.9($\pm0.7$)              \\
$\Omega_{in}$                     & 6.2428                      \\
$\Omega_{out}$                    & 5.6286                      \\
$\Omega_{1}=\Omega_{2}$           & 6.1118($\pm0.0474$)       \\
$T_{2}(K)$                        & 4993($\pm9$)                \\
$\Delta T(K)$                     & 237($\pm9$)  \\
$T_{2}/T_{1}$                     & 0.955($\pm0.002$)  \\
$L_{1}/(L_{1}+L_{2}$) (V)         & 0.347($\pm0.002$)           \\
$L_{1}/(L_{1}+L_{2}$) (R)         & 0.337($\pm0.002$)           \\
$L_{1}/(L_{1}+L_{2}$) (I)         & 0.330($\pm0.002$)          \\
$r_{1}(pole)$                     & 0.2857($\pm0.0014$)           \\
$r_{1}(side)$                     & 0.2991($\pm0.0016$)          \\
$r_{1}(back)$                     & 0.3391($\pm0.0026$)           \\
$r_{2}(pole)$                     & 0.4487($\pm0.0043$)            \\
$r_{2}(side)$                     & 0.4822($\pm0.0060$)           \\
$r_{2}(back)$                     & 0.5120($\pm0.0082$)           \\
$f$                               & $21\,\%$($\pm$8\,\%$$)    \\
$\Sigma{\omega(O-C)^2}$           & 0.00203                     \\
\hline
\end{tabular}
\end{center}
\end{table}

\begin{table}[!h]
\caption{Photometric solutions of UCAC4 436-062932 with O'Connell effect}\label{phsolutions}
\begin{center}
\begin{tabular}{lllllllll}
\hline
Parameters                        & V                            & $R_c$                        & $I_c$                    & Mean     \\
\hline
$g_{1}$                           & 0.32(fixed)                  & 0.32(fixed)                  & 0.32(fixed)              & 0.32(fixed)    \\
$g_{2}$                           & 0.32(fixed)                  & 0.32(fixed)                  & 0.32(fixed)              & 0.32(fixed)    \\
$A_{1}$                           & 0.50(fixed)                  & 0.50(fixed)                  & 0.50(fixed)              & 0.50(fixed)   \\
$A_{2}$                           & 0.50(fixed)                  & 0.50(fixed)                  & 0.50(fixed)              & 0.50(fixed)   \\
q ($M_2/M_1$ )                    & 2.65($\pm0.05$)              & 2.67($\pm0.04$)              & 2.81($\pm0.08$)          & 2.71($\pm0.03$) \\
$T_{1}(K)   $                     & 5230(fixed)                  & 5230(fixed)                  & 5230(fixed)              & 5230(fixed)  \\
$i(^{\circ})$                     & 83.3($\pm0.7$)               & 85.8($\pm0.9$)               & 84.7($\pm1.1$)           & 84.6($\pm0.53$)  \\
$\Omega_{in}$                     & 6.1485                       & 6.1755                       & 6.3635                   &           \\
$\Omega_{out}$                    & 5.5356                       & 5.5622                       & 5.7477                   &           \\
$\Omega_{1}=\Omega_{2}$           & 6.0257($\pm0.0629$)          & 6.0863($\pm0.0012$)          & 6.2128($\pm0.0993$)      & 6.1083($\pm0.0392$)         \\
$T_{2}(K)$                        & 5038($\pm38$)                & 4985($\pm15$)                & 4902($\pm23$)            & 4975($\pm16$)   \\
$\Delta T(K)$                     & 192($\pm38$)                 & 245($\pm15$)                 & 328($\pm23$)             & 255($\pm16$)  \\
$T_{2}/T_{1}$                     & 0.963($\pm0.007$)            & 0.953($\pm0.003$)            & 0.937($\pm0.004$)        & 0.951($\pm0.003$)  \\
$L_{1}/(L_{1}+L_{2}$)             & 0.340($\pm0.006$)            & 0.340($\pm0.002$)            & 0.341($\pm0.003$)        & 0.340($\pm0.002$)  \\
$r_{1}(pole)$                     & 0.2876($\pm0.0020$)          & 0.2843($\pm0.0015$)          & 0.2846($\pm0.0025$)      & 0.2855($\pm0.0012$)  \\
$r_{1}(side)$                     & 0.3011($\pm0.0023$)          & 0.2973($\pm0.0017$)          & 0.2981($\pm0.0028$)      & 0.2988($\pm0.0013$)   \\
$r_{1}(back)$                     & 0.3410($\pm0.0037$)          & 0.3351($\pm0.0025$)          & 0.3398($\pm0.0039$)      & 0.3386($\pm0.0020$)   \\
$r_{2}(pole)$                     & 0.4462($\pm0.0058$)          & 0.4448($\pm0.0050$)          & 0.4540($\pm0.0088$)      & 0.4483($\pm0.0039$)   \\
$r_{2}(side)$                     & 0.4791($\pm0.0079$)          & 0.4770($\pm0.0069$)          & 0.4889($\pm0.0123$)      & 0.4817($\pm0.0054$)  \\
$r_{2}(back)$                     & 0.5089($\pm0.0109$)          & 0.5059($\pm0.0094$)          & 0.5194($\pm0.0169$)      & 0.5114($\pm0.0074$)    \\
$f$                               & $20\,\%$($\pm$10\,\%$$)      & $14\,\%$($\pm$9\,\%$$)       & $25\,\%$($\pm$16\,\%$$)  & $20\,\%$($\pm$7\,\%$$)   \\
$\theta(^{\circ})$                & 146($\pm37$)                 & 102(fixed)                   & 102($\pm72$)             & 117($\pm27$)       \\
$\psi(^{\circ})$                  & 300($\pm18$)                 & 279($\pm5$)                  & 265($\pm4$)              & 281($\pm6$)      \\
$r$(rad)                          & 0.3063($\pm0.2000$)          & 0.2433($\pm0.0151$)          & 0.3322($\pm0.0574$)      & 0.2939($\pm0.0695$)      \\
$T_f$                             & 0.82(fixed)                  & 0.82(fixed)                  & 0.82(fixed)              & 0.82(fixed)      \\
$\Sigma{\omega(O-C)^2}$           & 0.00085                      & 0.00060                      & 0.00032                  &     \\
\hline
\end{tabular}
\end{center}
\end{table}

\begin{figure}[!h]
\begin{center}
\includegraphics[width=14cm]{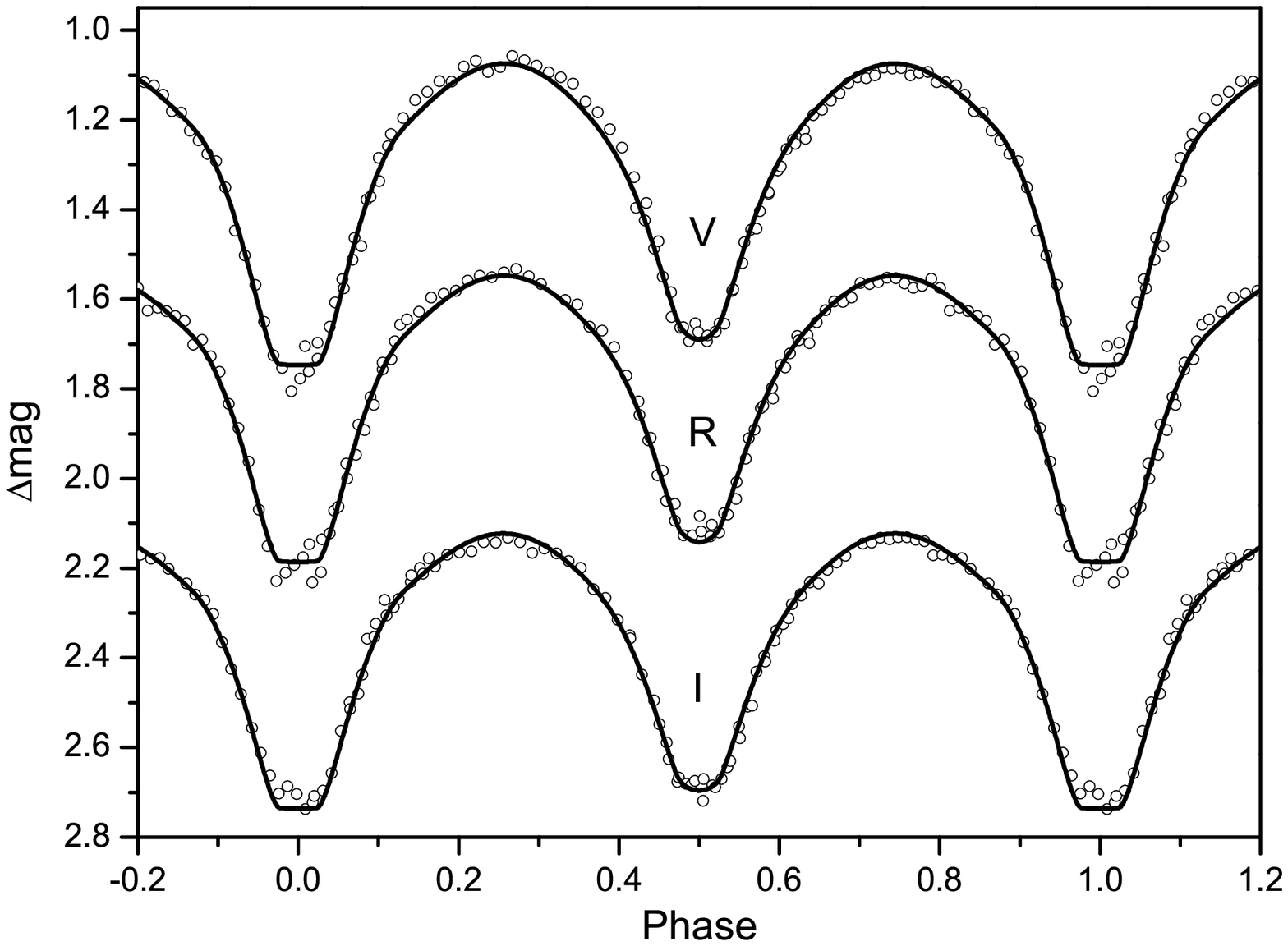}
\caption{Observed (open circles) and theoretical (solid lines) light curves in the $V R_c$ and $I_c$ bands of UCAC4 436-062932 without O'Connell effect. The standard deviation of the fitting residuals is 0.022 mag for $V$ band, 0.020 mag for $R_c$ band and 0.017 mag for $I_c$ band, respectively.}
\end{center}
\end{figure}

\begin{figure}[!h]
\begin{center}
\includegraphics[width=14cm]{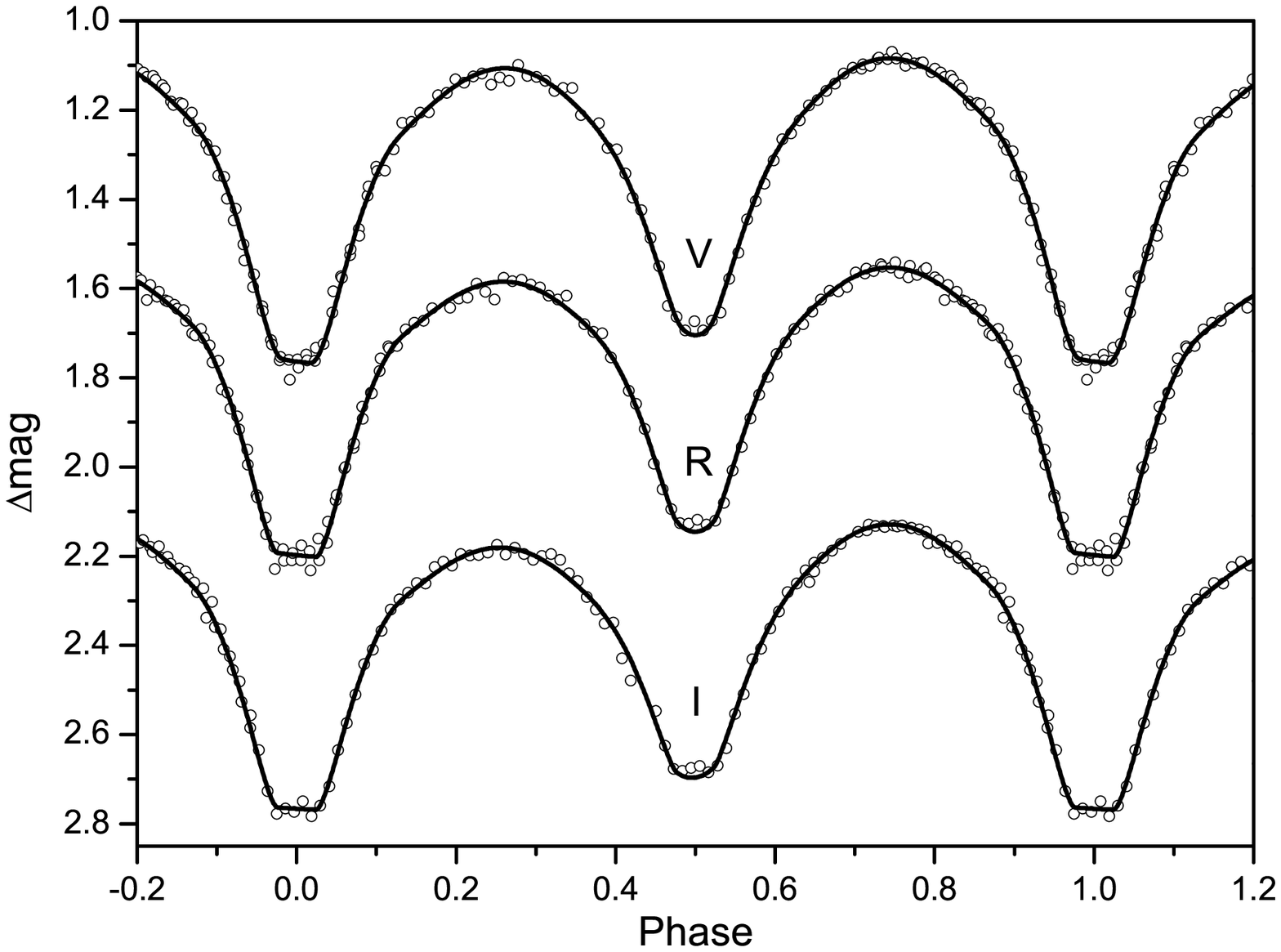}
\caption{Observed (open circles) and theoretical (solid lines) light curves in the $V R_c$ and $I_c$ bands of UCAC4 436-062932 with O'Connell effect. The standard deviation of the fitting residuals is 0.015 mag for $V$ band, 0.014 mag for $R_c$ band and 0.014 mag for $I_c$ band, respectively.}
\end{center}
\end{figure}

\begin{figure}[!h]
\begin{center}
\includegraphics[width=14cm]{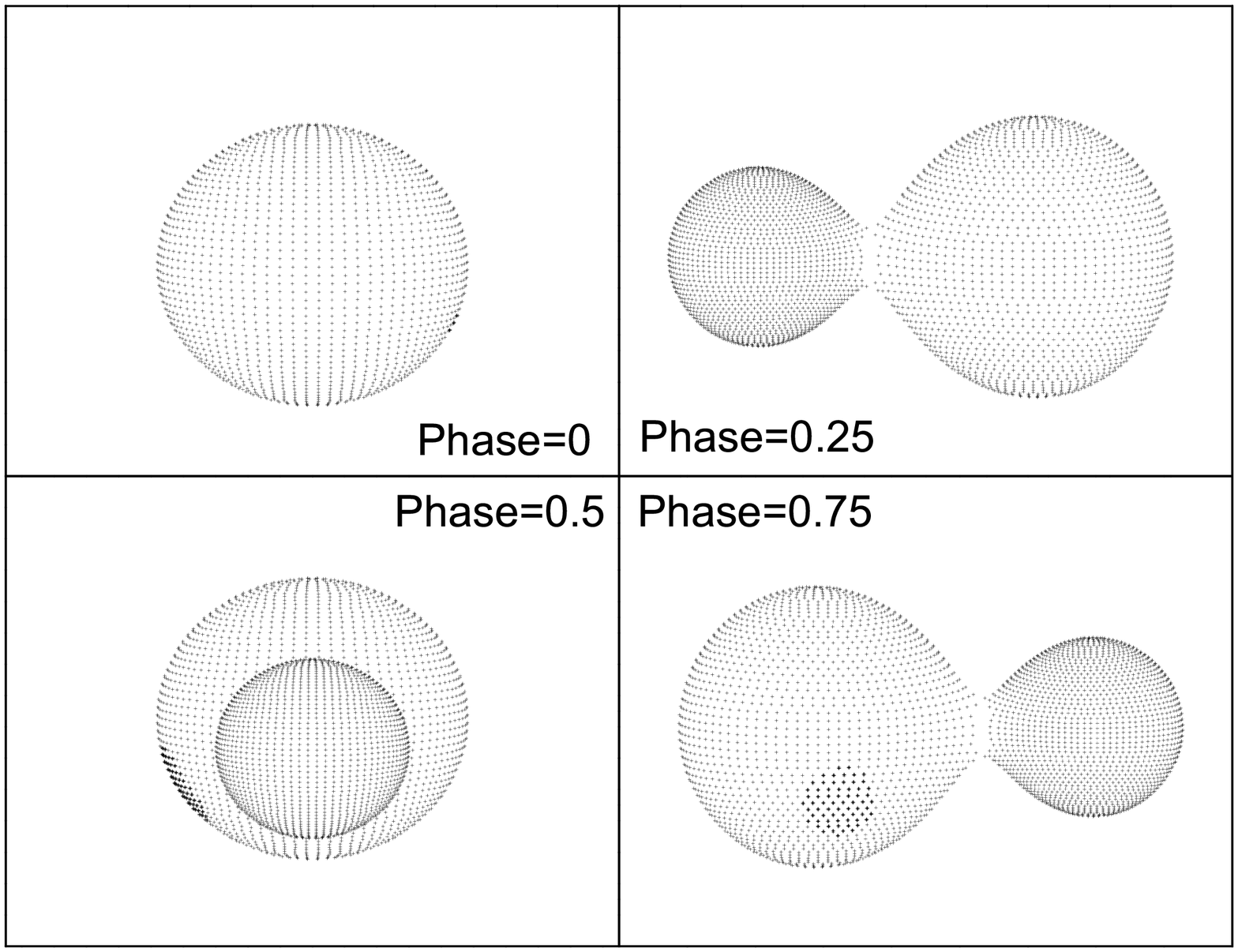}
\caption{Contact configurations of UCAC4 436-062932 with spot at phase 0.0, 0.25, 0.5 and 0.75.}
\end{center}
\end{figure}

\section{Discussions and Conclusions}
The two sets of multi-color light curves give nearly consistent results, indicating that UCAC4 436-062932 is a shallow contact binary system with a low degree of contact (about $f = 20\,\%$). The derived mass ratio $q = M_2/M_1 = 2.7$ implies that the UCAC4 436-062932 is a W-subtype binary system. The less massive component is about $250K$ hotter than the more massive one. The extremely high orbital inclination (about $i = 85^{\circ}$) suggests that the eclipse during the primary minimum is a total eclipse and the obtained parameters are very reliable.

In our two sets of light curves, O'Connell effect appears in the light curves observed on 2014 June 18 and June 19. However, the light curves observed on 2014 May 23 and June 19 have almost equal brightness at the two maxima. We try to model the asymmetric light curves (with O'Connell effect) with spot model, and it really gives a quite satisfying results. Circumstellar matter model suggested by \citet{2003ChJAA...3..142L} and some other models can also give quite well fitting results and explanations to the asymmetric light curves. As considering the two sets of light curves are observed in such a short time (cover about 30 days), the spot model may be more appropriate in the case of UCAC4 436-062932 for the lifetime of a star spot can be just several ten days. It is not possible that the effect of circumstellar matter and Coriolis force would appear or disappear in a month. It means that the star spot is not exist or too small to be considered when UCAC4 436-062932 was observed on 2014 May 23. The star spot appeared when we observed the light curves of UCAC4 436-062932 on 2014 June 18, thus we obtained light curves with O'Connell effect. However, we can't simply conclude that spot model is the only explanatory mechanism for O'Connell effect. In other binary systems, O'Connell effect may be mainly dominated by circumstellar matter or Coriolis force. To conform this, more photometric observations and analysis of light curves with and without O'Connell effect are needed in the future.

\section{Acknowledgments}
This work is supported by the Chinese Natural Science Foundation (Grant No. 11133007, 11325315 and 11203066), the Strategic Priority Research Program ``The Emergence of Cosmological Structure'' of the Chinese Academy of Sciences (Grant No. XDB09010202) and the Science Foundation of Yunnan Province (Grant No. 2012HC011). This research has made use of the fourth US Naval Observatory CCD Astrograph Catalog (UCAC4).

\end{document}